\documentclass[aps,prr,twocolumn,floatfix,amsmath,amssymb,secnumarabic,]{revtex4-2} % change prl \to prr for referencing Sections

\usepackage{graphicx}
\usepackage{braket}
\usepackage{color}
\usepackage[hidelinks]{hyperref}
\usepackage{orcidlink}
\usepackage{dsfont}
\usepackage{dcolumn}

\hypersetup{colorlinks=true, linkcolor=blue, citecolor=blue, urlcolor=blue}

\usepackage[utf8]{inputenc}

\definecolor{newgreen}{rgb}{0,0.5,0}

\begin{document}

\title{The Cumulant Expansion Approach: The Good, The Bad and The Ugly}

\author{Johannes Kerber\, \orcidlink{0009-0002-1957-8008}}
\email[]{johannes.kerber@uibk.ac.at}
\author{Helmut Ritsch\,\orcidlink{0000-0001-7013-5208}}
\author{Laurin Ostermann\,\orcidlink{0000-0001-8508-9785}}
\email[]{laurin.ostermann@uibk.ac.at}

\affiliation{Institut f\"ur Theoretische Physik, Universit\"at Innsbruck\\Technikerstra{\ss}e\,21a, A-6020 Innsbruck, Austria} 

\begin{abstract}
The configuration space, i.e.\ the Hilbert space, of compound quantum systems grows exponentially with the number of its subsystems: its dimensionality is given by the product of the dimensions of its constituents. Therefore a full quantum treatment is rarely possible analytically and can be carried out numerically for fairly small systems only. Fortunately, in order to obtain interesting physics, approximations often very well suffice. One of these approximations is given by the cumulant expansion, where expectation values of products of operators are approximated by products of expectation values of said operators, neglecting higher-order correlations. The lowest order of this approximation is widely known as the mean field approximation and used routinely throughout quantum physics. Despite its ubiquitous presence, a general criterion for applicability and  convergence properties of higher order cumulant expansions remains to be found. In this paper, we discuss two problems in quantum electrodynamics and quantum information, namely the collective radiative dissipation of a dipole-dipole interacting chain of atoms and the factorization of a bi-prime by annealing in an adiabatic quantum simulator. In the first case we find smooth, convergence behavior, where the approximation performs increasingly better with higher orders, while in the latter going beyond mean field turns out useless and, even for small system sizes, we are puzzled by numerically challenging and partly non-physical solutions.
\end{abstract}
\date{\today}
\maketitle

\section{Introduction}
The mathematical framework used for quantum systems consists of a square-integrable vector space, the so-called Hilbert space, which supports state vectors and linear mappings, i.e.\ operators representing actions on these state vectors. The dimension of this Hilbert space is given by the number of basis states, where joint systems feature a dimension that is the product of the dimensions of its constituents creating an exponential scaling of the configuration space with the number of components. Therefore, 'exact' calculations of larger systems become 'exponentially more difficult', both, analytically as well as in numerical simulations. Consequently, an entire zoo of approximation techniques has been developed in order to obtain physically valuable and useful results~\cite{dum1992monte,herman1994dynamics,sinatra2002truncated,cirac2021matrix,verstraete2023density}.

\begin{figure}[h!]
\centering
\includegraphics[clip, trim=2.5cm 18.9cm 2.5cm 1.5cm, width=.55\textwidth]{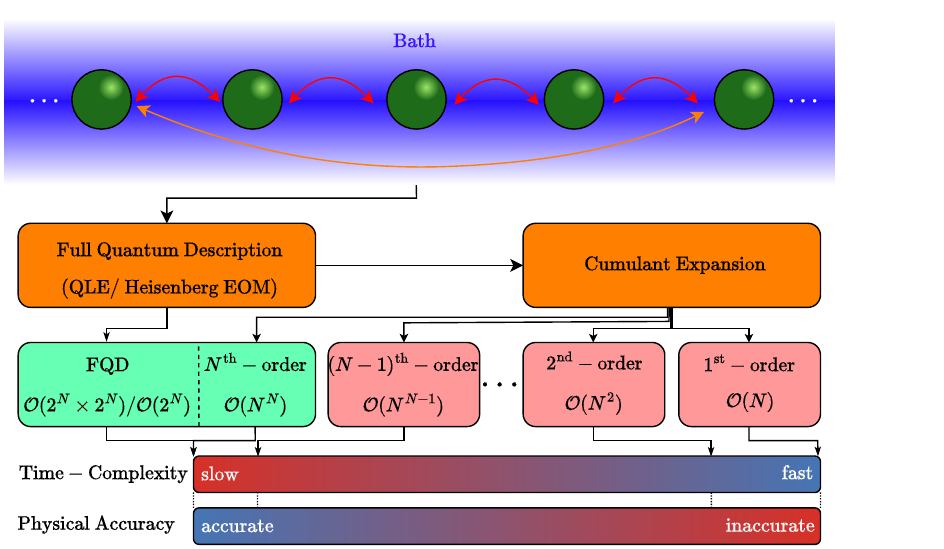}
\caption{\emph{Model Systems and Order of Expansion.} Our toy models consist of $N$ interacting identical two-level systems. The dimension of a full quantum treatment scales exponentially with $N$ as $2^N \times 2^N$  (Quantum Langevin Equation, QLE). It is commonly assumed and remains to be proven rigorously, that the cumulant expansion approximates the solution of the QLE better and better by increasing its order $o$. The first order of the cumulant expansion, dubbed mean field approximation, scales only linearly with system size, i.e.\ $\mathcal{O}(N)$, but does not account for quantum correlations. Including all pairwise quantum correlations in second order gives $\mathcal{O}(N^2)$. Higher order cumulants account for multi-particle quantum correlations at the cost of a polynomial increase in the number of equations, i.e.\ $\mathcal{O}(N^o)$ as a function of $o$. By looking at the scaling of the equations, one can find $o < N\ln(4)/\ln(N)$, identifying the highest order where the cumulant expansion is computationally faster than the full quantum treatment. The schematics visualizes the trade-off between time-complexity and physical accuracy upon increasing the order of the expansion.}
\label{fig.1}
\end{figure}

A longstanding successful technique and the matter of interest of the present manuscript is the cumulant expansion method~\cite{kubo1962generalized,van1974cumulant}, where the full dynamics of the system given by the Schrödinger or the master equation is transformed into the Heisenberg picture, i.e.\  the time evolution is transferred from the state vector or density matrix to the observables. Subsequently, a (possibly open) set of differential equations for the operators can be derived, where the time dependence of the operators is tied to higher-order products of other operators. Upon taking expectation values of these equations one can obtain an approximated closed set of differential equations by replacing the expectation values of products of operators of higher order with products of expectation values of operators of lower order. This implicitly assumes statistical independence of higher order correlations. The general procedure of the cumulant expansion method is depicted in Fig.~\ref{fig.1}.

Applying the first-order of the cumulant expansion method to a given problem is known as mean field approximation~\cite{kolomietz2020mean}, neglecting all quantum correlations between operators. In analogy to the Ehrenfest theorem, where the equations of motion of the expectation values of position and momentum of a quantum harmonic oscillator correspond to the Euler-Lagrange equations of its classical analogue~\cite{ehrenfest1927bemerkung}, one interprets this first order approximation as close to the 'classical' limit, ignoring key quantum correlation effects. Including all pairwise quantum correlations typically already accounts for a large part of the quantum fluctuations in the system, e.g., it includes equations for operator uncertainty values. Many successful applications as, e.g., in laser theory are based on second order cumulant expansions \cite{risken1996statistical,henschel2010cavity,bychek2023superradiant}. However, one has to keep in mind, that the solutions do not guarantee a consistent physical time evolution corresponding to a  physically allowed quantum state or density matrix.

Now, given  the cascaded nature of the underlying commutators that have to be evaluated in order to go to higher orders, the expansion is seldomly performed beyond mean field or second order~\cite{cirac2021matrix, rubies2023characterizing, henschel2010cavity}. Up until very recently, although the general recipe is straight-forward, the cumbersome algebraic book keeping has outweighed the prospective benefits of performing these calculations. With the advent of symbolic frameworks~\cite{Gowda2022High,wester1999computer} and, in particular, our own QuantumCumulants.jl toolbox~\cite{plankensteiner2022quantumcumulants}, cumulant expansion equations can now be, at least in principle, derived automatically up to an arbitrary order. The toolbox currently allows for finite and countably infinite Hilbert spaces. The simplest of those types of systems (and also the subject of the present study) is a compound system of $N$ identical two-level systems. As the square of any combination of single constituent operators is proportional to the identity, an $N$-th order cumulant expansion is identical to the full quantum dynamics (FQD) (see Fig.~\ref{fig.1}).

At this point no general applicability criterion for the cumulant expansion exists nor are there general proofs of convergence. The usual and widely accepted procedure goes like this: solve a small exactly treatable toy system for its full dynamics, solve the same system in the desired cumulant expansion approximation order and do this for increasing system size usually in steps of $1$ until the exact solution becomes impossible to calculate. If the targeted expectation values of the full solution vs.\ the approximation agree on a reasonable level, deem the approximation legit and scale up the system for the approximation. While this certainly has its legitimacy and has been applied successfully many times~\cite{risken1996statistical,zhuravlev2020cumulant,drut2016entanglement} including our own work~\cite{kramer2016optimized,fasser2024subradiance}, the convergence of the method is by no means rigorous or strictly quantifiable.

In this paper we intend to show that this usual procedure can work very well ("The Good"), fail ("The Bad") and leave one completely puzzled with chaotic uncontrollable solutions ("The Ugly"), even for surprisingly small size examples. To this end we look at two apparently quite similar toy setups, a handful of two-level systems in two scenarios: dipole-dipole mediated collective radiation ("The Good") and a simple bi-prime factorization algorithm by means of adiabatic quantum computing ("The Bad" and "The Ugly"). The observations in this manuscript shall not only motivate the search for a general and universal criterion for the applicability and quality of the cumulant expansion  but also intend to demonstrate the underlying complexity. Thus, we solely compare the CEM to the full quantum description of specific physical quantities of the fundamental model rather than physical effects such as, e.g.,\ super- and subradiance or an optimization of the factorization Hamiltonian. Before we begin, let us briefly recap the concept of the cumulant expansion method (CEM) and its mathematical underpinnings.

\section{The Cumulant Expansion}
Let $X_1,X_2,\cdots,X_n$ be $n$ ordered random variables. The joint cumulant of these variables is defined as the coefficient of the Maclaurin series of the so-called multivariate cumulant generating function \cite{leonov1959method}. The joint cumulant $\langle \cdot \rangle_{\text{c}}$ can be written combinatorially as~\cite{kubo1962generalized}

\begin{equation}
    \langle X_1 X_2 \cdots X_n \rangle_{\text{c}} = \sum_{p \in \mathcal{P}(\mathcal{I})} (p-1)!(-1)^{|p|-1}\prod_{B\in p}\left\langle \prod_{i\in B}X_i \right\rangle,\label{eq.joint}
\end{equation}
where $\mathcal{I} = \{1,2,\cdots,n\}$ denotes the index set with all possible partitions $\mathcal{P}(\mathcal{I})$ and $\langle \cdot \rangle$ is the corresponding expectation value. If the random variables $X_1,X_2,\cdots,X_n$ (or subsets of them) are statistically independent, the joint cumulant fulfills $\langle X_1 X_2 \cdots X_n \rangle_c = 0$. This allows to rearrange Eq.~(\ref{eq.joint}) to

\begin{equation}
        \langle X_1 X_2 \cdots X_n \rangle = \sum_{p \in \mathcal{P}(\mathcal{I})\backslash\mathcal{I}} (p-1)!(-1)^{|p|}\prod_{B\in p}\left\langle \prod_{i\in B}X_i \right\rangle.\label{eq.expan}
\end{equation}
One recognizes the beneficial form in Eq.~(\ref{eq.expan}), since it provides a recursive expression of $\langle X_1 X_2 \cdots X_n \rangle$ in terms of a sum of products of expectation values up to order $n-1$, i.e., $\langle\prod_{i\in B}X_i \rangle$, where $|i|$ is the corresponding order. In quantum mechanics, observables are used rather than random variables. In \cite{kubo1962generalized}, the CEM is also generalized from random variables $X_1,X_2,\cdots,X_n$ to observables $\hat{X}_1,\hat{X}_2,\cdots,\hat{X}_n$. Hence, we can apply this method to expectation values $\langle \hat{X}_1 \hat{X}_2 \cdots \hat{X}_n \rangle$. These expectation values embody quantum correlations of $n$-th order. Thus, the CEM allows to describe a quantum system classically, neglecting all quantum correlations (e.g.\ Ehrenfest theorem for a harmonic oscillator where it corresponds to the classical equations of motion), partially quantum-mechanically by taking correlations of higher order into account or even fully quantum mechanically in the special case of a finite-dimensional Hilbert space. Here, the order of the approximation matches the number of subsystems.

\section{Models}
\subsection{Chain of Dipole-Dipole Interacting Two-Level Quantum Emitters}\label{suse.mod1}
\begin{figure}[h!]
    \centering
        \includegraphics[clip, trim=4.5cm 22.5cm 1.5cm 2cm, width=.70\textwidth]{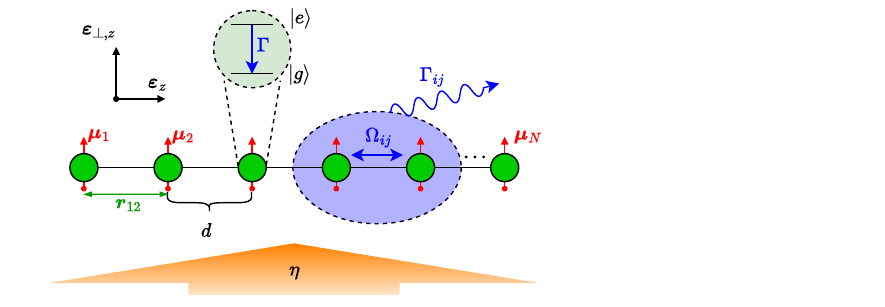}
    \caption{Chain of dipole-dipole interacting two-level QEs equidistantly separated by $d$ and constantly driven via $\eta$. The dipole moments $\boldsymbol{\mu}_i$ of all QEs are aligned in the same direction (red arrows). We choose $\boldsymbol{\mu}_i = \mu_i\boldsymbol{\varepsilon}_{\perp,z}$ to be perpendicular to the distance vector $\boldsymbol{r}_{ij} = \boldsymbol{r}_i - \boldsymbol{r}_j = d(i-j)\boldsymbol{\varepsilon}_{z}$ between the $i$-th and $j$-th QE (green arrow).}
    \label{fig.chain}
\end{figure}

We consider an equidistant chain of $N$ identical two-level quantum emitters (QEs) coupled via dipole-dipole interaction in free space and driven by a constant coherent drive (see Fig.~\ref{fig.chain}).

The two single-emitter states $\ket{e}$ and $\ket{g}$ are separated by $\omega_0$ ($\hbar = 1$) and spontaneous emission from the excited state to the ground state occurs with a rate of $\Gamma$. The dipole-dipole interaction induces collective dispersion as well as collective decay~\cite{lehmberg1970radiation}. The Hamiltonian of this open quantum system is given by
\begin{equation}
\hat{H} = \omega_0\sum_{i}\hat{\sigma}^+_i\hat{\sigma}^-_i + \sum_{i \not = j} \Omega_{ij} \hat{\sigma}^+_i \hat{\sigma}^-_j + \eta\sum_i(\hat{\sigma}_i^{+} + \text{h.c.}),\label{eq:ham1}
\end{equation}
where $\hat{\sigma}^+_i~(\hat{\sigma}^-_i)$ refers to the raising (lowering) operator of the $i$-th QE, $\Omega_{ij}$ define the coherent couplings between the $i$-th and $j$-th QE and $\eta\in\mathbb{R}$ represents the constant drive strength of the system. The collective dissipation is modeled by a Liouvillian super-operator
\begin{equation}
\hat{\mathcal{L}} \left[ \hat{\rho} \right] = \frac{1}{2} \sum_{i, j} \Gamma_{ij} \left( 2 \hat{\sigma}^-_i \hat{\rho} \hat{\sigma}^+_j - \hat{\sigma}^+_i \hat{\sigma}^-_j \hat{\rho} - \hat{\rho} \hat{\sigma}^+_i \hat{\sigma}^-_j \right),
\end{equation}
where $\hat{\rho}$ denotes the density operator and $\Gamma_{ij}$ are the incoherent couplings with $\Gamma_{ii} = \Gamma$. The exact distant-dependent behaviors of $\Omega_{ij} = \Omega_{ij}(k_0r_{ij})$ and $\Gamma_{ij} = \Gamma_{ij}(k_0r_{ij})$, respectively, are adapted from \cite{ostermann2016collective}. Here, $k_0 = 2\pi/\lambda$ states the wave vector and $r_{ij} = d\cdot|i - j|$ is the distance between the $i$-th and $j$-th QE. Choosing the alignments of $\boldsymbol{\mu}_i$ and $\boldsymbol{r}_{i,j}$ as in Fig.~\ref{fig.chain} yields the simplified expressions:

\begin{align}
\Omega_{ij} &= \frac{-3\Gamma}{4}\bigg(\frac{\cos{(\xi_{ij})}}{\xi_{ij}} - \bigg( \frac{\sin{(\xi_{ij})}}{\xi_{ij}^2} + \frac{\cos{(\xi_{ij})}}{\xi_{ij}^3}\bigg)\bigg),\\
\Gamma_{ij} &= \frac{3\Gamma}{2}\bigg(\frac{\sin{(\xi_{ij})}}{\xi_{ij}} + \bigg( \frac{\cos{(\xi_{ij})}}{\xi_{ij}^2} - \frac{\sin{(\xi_{ij})}}{\xi_{ij}^3}\bigg)\bigg),
\end{align}
where $\xi_{ij} = k_0r_{ij}$. In the full quantum treatment, the system's dynamics are governed by the Lindblad master equation

\begin{equation}
\partial_t \hat{\rho} = i [ \hat{\rho}, \hat{H} ] + \hat{\mathcal{L}}[\hat{\rho}],\label{eq.master}
\end{equation}
which, written in matrix representation, features a size of $4^N \times 4^N$, which ostensively scales exponentially with the number of emitters $N$. Hence, for numerically treating larger finite systems, one needs to turn to approximation methods in order to reduce the number of equations that need to be solved. We want to use the CEM, therefore, we introduce the expectation value $\langle \hat{\mathcal{O}} \rangle = \text{tr}{(\hat{\mathcal{O}}\hat{\rho})}$ of an arbitrary system operator $\hat{\mathcal{O}}$ which yields the expression

\begin{equation}
    \partial_t \langle\hat{\mathcal{O}}\rangle = -i \langle[ \hat{\mathcal{O}}, \hat{H} ]\rangle + \frac{1}{2} \sum_{i, j} \Gamma_{ij} \left( 2 \langle\hat{\sigma}^-_i \hat{\mathcal{O}} \hat{\sigma}^+_j\rangle - \langle\{\hat{\sigma}^+_i \hat{\sigma}^-_j, \hat{\mathcal{O}}\}\rangle\right),
\end{equation}
where we use the Quantum Langevin Equation (QLE) \cite{Gardiner1985}:

\begin{align}
    \partial_t\hat{\mathcal{O}} = &-i[ \hat{\mathcal{O}}, \hat{H} ]\nonumber\\
    &- \sum_{i,j}\frac{\Gamma_{ij}}{2}([\hat{\mathcal{O}},\hat{\sigma}^+_i]\hat{\sigma}_j^- + \hat{\sigma}_i^+[\hat{\mathcal{O}},\hat{\sigma}^-_j]) \nonumber\\
    &- \sum_{i,\mu}([\hat{\mathcal{O}},\hat{\sigma}^+_i]C_{i\mu}\hat{b}_{\mu,\text{in}} + \hat{b}_{\mu,\text{in}}^{\dagger}C_{i\mu}^*[\hat{\mathcal{O}},\hat{\sigma}^-_i]).
\end{align}

Here, $\Gamma_{ij} = \sum_{\mu}C_{i\mu}C_{j\mu}^*$ are the incoherent decay rates with decay rate coefficients $C_{i\mu}$, $\hat{b}_{\mu,\text{in}}$ is the input noise operator associated with bath channel $\mu$, fulfilling $[\hat{b}_{\mu,\text{in}}(t),\hat{b}_{\mu',\text{in}}^{\dagger}(t')] ~= \delta_{\mu,\mu'}\delta(t-t').$ We assume white noise, i.e.\ $\langle \hat{b}_{\mu,\text{in}}(t) \rangle = 0$ and $\langle \hat{b}_{\mu,\text{in}}(t)\hat{b}_{\mu',\text{in}}(t') \rangle = \delta_{\mu,\mu'}\delta(t-t')$. This implies the cancellation of all terms containing quantum noise operators. The brackets $[\cdot,\cdot]$ and $\{\cdot,\cdot\}$ represent the commutator and anti-commutator, respectively. The CEM allows the description of the problem's dynamics by a system of ordinary differential equations (ODEs) with reduced number of equations.

\subsection{Bi-Prime Factorization Problem}\label{suse.mod2}

\begin{figure}[h!]
    \centering
        \includegraphics[clip, trim=3.5cm 14.3cm 3.0cm 10.0cm, width=.65\textwidth]{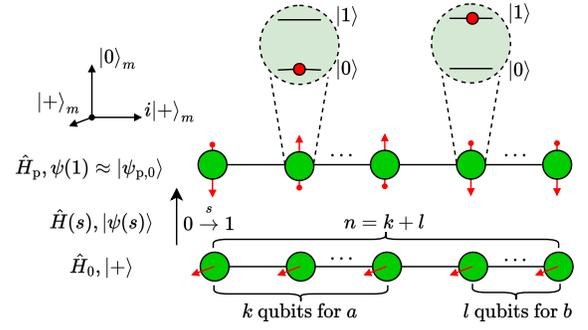}
    \caption{Adiabatic quantum algorithm for the bi-prime factorization problem. The red arrows represent the Bloch vector alignments of the corresponding qubits. The visualization of each vector is done by utilizing the Bloch representation $\{\ket{+}_m = 1/\sqrt{2}(\ket{0}_m + \ket{1}_m),i\ket{+}_m=1/\sqrt{2}(\ket{0}_m + i\ket{1}_m),\ket{0}_m\}\in\mathcal{H}_m\cong\mathbb{C}^2$ where the full quantum state of the system is a vector within the collective Hilbert space $\ket{\psi(s)} \in\mathcal{H}$. The lower $n = k+l$ qubit arrangement represents the ground state of $\hat{H}_0$ at $s = 0$. Turning on the adiabatic process, i.e.\ $s = 0\to1$, changes the system's internal Bloch vector alignments which leaves (ideally) an array of 'up'- or 'down'-pointing vectors (upper vector arrangement). Measuring the projectors $\hat{a}_i,\hat{b}_j$ results in a sequence of the classical bit-digits, fulfilling the bi-prime problem $\omega=ab.$}
    \label{fig.biprime}
\end{figure}

A given bi-prime $\omega\in\mathbb{N}_{>1}$ fulfills the relation $\omega = a\cdot b$, where $a,b$ are prime numbers. If $\omega$ is chosen under certain conditions\cite{helfgott2020improved,lehman1974factoring,mckee1996turning}, the determination of the factors utilizing a classical computer is impractical, since all known classical factoring algorithms scale exponentially in time-complexity \cite{lehman1974factoring}. However, the harness of quantum theory allows the construction of quantum algorithms which can solve the problem in polynomial time-complexity \cite{shor1999polynomial}. In our case, we consider the factorization problem as Quantum Annealing Problem (QAP). First, we describe the factorization as a classical optimization problem (COP). The factors $a,b$ can be determined by minimizing the cost function $f_{\omega}(x,y) = (\omega - xy)^2$. We can express $f_{\omega}(x,y)$ in terms of binary representation sums \cite{peng2008quantum,schaller2007}:
\begin{equation}
    f_{\omega}(a_0,a_1,\cdots,b_{l-1},b_l) = \Big(\omega - \sum_{i = 0}^{k}\sum_{j = 0}^{l}2^{i+j}a_ib_j\Big)^2,\label{eq.cost}
\end{equation}
where we use $a,b$ rather than $x,y$ and $k,l\in\mathbb{N}$. The factors $a$ and $b$ possess $k+1$ and $l+1$ classical bit digits, respectively. For simplicity, we only consider odd $\omega$ which implies $a_0 = 1 = b_0$. This leaves us with $n = k + l$ unknown classical bits which need to be determined to solve the bi-prime problem. Next, we treat the COP as Quantum Optimization Problem (QOP). Since the classical bit digits are either $0$ or $1$, we consider generic qubits rather than two-level emitters, omitting all forms of interaction or decay. We use the notation $\ket{0} = \ket{g}$, $\ket{1} = \ket{e}$ and define the number of qubits as $n = N$. Then, we find an arbitrary state $\ket{\psi} \in \mathcal{H} = \bigotimes_{m = 1}^{n} \mathcal{H}_m \cong \left(\mathbb{C}^2\right)^{\otimes n}$ of shape $\ket{\psi} = \bigotimes_{i = 1}^{k}\ket{a_i}_i\otimes\bigotimes_{j = 1}^{l}\ket{b_j}_{k+j}$,
where $\ket{a_i}_i\in\mathcal{H}_i$ and $\ket{b_j}_{k+j}\in\mathcal{H}_{k+j}$. We notice that the projectors $\hat{\sigma}_m^{22} = 1/2\cdot(\mathds{1} - \hat{\sigma}^{z}_m)$ fulfill the eigenvalue equations $\hat{\sigma}_m^{22}\ket{z_m}_m = z_m\ket{z_m}_m$, where $z_m\in\{0,1\}$ and $\mathds{1}$ is the unit matrix. This allows to express Eq.~(\ref{eq.cost}) as a QOP in form of energy minimization. We then find the problem Hamiltonian \cite{peng2008quantum,schaller2007,yang2025improving} 
\begin{equation}
    \hat{H}_{\text{p}} = \hbar\Omega\Big(\omega\mathds{1} - \Big(\mathds{1} + \sum_{i = 1}^{k}2^{i}\hat{a}_i\Big)\Big(\mathds{1} + \sum_{j = 1}^{l}2^{j}\hat{b}_j\Big)\Big)^2,\label{eq.Hp}
\end{equation}
where $\hbar\Omega = 1s^{-1}$, $\hat{a}_i = \hat{\sigma}_i^{22}$ and $\hat{b}_j = \hat{\sigma}_{k+j}^{22}$. We recognize, that there are two possible solutions for the ground state energy ($a\cdot b = b\cdot a$) and assume for further discussion $k \geq \lceil n/2 \rceil > l.$ With increasing $n$ the problem's time-complexity grows exponentially, implying the diagonalization of $\hat{H}_{\text{p}}$ for determining the ground state $\ket{\psi_{\text{p},0}}$, which encodes the solution of the bi-prime problem, becomes computationally more expensive. The QAP overcomes this problem by utilizing the adiabatic Hamiltonian $\hat{H}(s) = (1 - s)\hat{H}_0 + s\hat{H}_{\text{p}}$, where $s = t/T$ with $t\in[0,T]$ and $T\in\mathbb{R}_+$ denotes the overall-evolution time of the adiabatic process \cite{born1928beweis,farhi2000quantum,messiah2014quantum}. We choose an initial Hamiltonian $\hat{H}_0$ which possesses an easy ground state $\ket{\psi_{\text{0},0}} = \ket{+} = 1/\sqrt{2^n}\bigotimes_{m = 1}^{n}(\ket{0}_m + \ket{1}_m)$. The corresponding Hamiltonian is given by

\begin{equation}
    \hat{H}_0 = -\xi\sum_{m = 1}^{n}\hat{\sigma}^{x}_m,
\end{equation}
where $\xi\in\mathbb{R}$. According to the adiabatic theorem \cite{born1928beweis,messiah2014quantum}, if $T$ is large enough and there exists a non-zero energy gap between ground state energy and the rest of the discrete spectrum of $\hat{H},~\forall s\in[0,1]$, then, we find a state $\ket{\psi(s)}\in\mathcal{H}$ fulfilling $\ket{\psi(s = 0)} = \ket{\psi_{\text{0},0}} \rightarrow \ket{\psi(s = 1)} \approx \ket{\psi_{\text{p},0}}$. During the time-evolution we force the qubits into a position aligned either in $\ket{0}_m$ or $\ket{1}_m$ direction. We measure the projectors $\hat{a}_i,\hat{b}_j$ after the evolution, i.e.\ at $s = 1$, which yields $\bra{\psi(1)} \hat{a}_i \ket{\psi(1)}  \approx a_i$ and $\bra{\psi(1)} \hat{b}_j \ket{\psi(1)} \approx b_j$, respectively. Thus, the measurements of the projectors correspond to the classical bit digits of $a,b$. The setup is depicted in Fig.~\ref{fig.biprime}.

Let us point out that the adiabatic quantum algorithm discussed in this section might not be the most efficient way to solve the bi-prime factorization problem (four-body interaction). Using a different algorithm might provide different CEM results. For instance, the algorithm investigated in \cite{schaller2007} (reduction to two-body interaction by introducing auxiliary qubits) could resolve the CEM discrepancies, discussed in Sec.~\ref{suse:thebad} and in Sec.~\ref{suse:theugly}, for this particular problem. However, since we want to present a case where the CEM fails even for a fairly small system size, we deliberately chose not to employ optimized factorization Hamiltonians in this work.

The FQD of an arbitrary system operator $\hat{\mathcal{O}}$ (not specifically time-dependent in the Schr\"odinger picture) for this closed quantum system is governed by the Heisenberg equation of motion (EOM):

\begin{align}
    \partial_t\hat{\mathcal{O}} = i[\hat{H},\hat{\mathcal{O}}].\label{eq.Heisenberg}
\end{align}
We make use of the generalized Ehrenfest theorem \cite{ehrenfest1927bemerkung}:

\begin{align}
    \partial_t\langle\hat{\mathcal{O}}\rangle = i\langle[\hat{H},\hat{\mathcal{O}}]\rangle,
\end{align}
which allows to construct once again an ODE system with, applying the CEM, reduced number of equations. 

\subsection{Measure of Error}
We want to compare the solutions of the CEM in various orders with the results of the full quantum treatment (master equation in Eq.~(\ref{eq.master})/ Heisenberg EOM in Eq.~(\ref{eq.Heisenberg})). Therefore, we define the general error measure

\begin{align}
    \tilde{\Delta}_{o,\tilde{k},(m)} = 
    \begin{cases}
    \sum_{i = 1}^{M}\Delta_{o,\tilde{k}}(\chi_i),\text{~model~1} \\
    \sum_{i = 1}^{M}\Delta_{o,\tilde{k},m}(\chi_i), \text{~model~2}
    \end{cases}\label{eq:error}
\end{align}
where $M$ is the number of iteration points specified to each of the two problems, $o$ denotes the order of the CEM, $\tilde{k}\in\{22,z,x\}$ states the nature of a given system operator, e.g., $\tilde{k} = 22\to\hat{\sigma}^{22}$, and $\chi_i\in\{\Gamma t_i,s_i\}$ is the corresponding time of the regarded system. The index $m$ remains for model 2, because we do not regard the mean expectation values. We define the squared differences (SD)

\begin{align}
    \Delta_{o,\tilde{k},(m)}(\chi_i) =  
    \begin{cases}
    (\langle\hat{\sigma}^{\tilde{k}}(\chi_i)\rangle_o - \langle \hat{\sigma}^{\tilde{k}}(\chi_i) \rangle_{\textbf{FQD}})^2,\text{~model~1}\\
    (\langle\hat{\sigma}_m^{\tilde{k}}(\chi_i)\rangle_o - \langle \hat{\sigma}_m^{\tilde{k}}(\chi_i) \rangle_{\textbf{FQD}})^2,\text{~model~2}
    \end{cases}
    \label{eq:error1}
\end{align}
where $\langle\hat{\sigma}^{\tilde{k}}(\chi_i)\rangle_o = 1/N\cdot\sum_{m = 1}^N\langle\hat{\sigma}^{\tilde{k}}_m(\chi_i)\rangle_o$ is the time-evolving (given by $\chi_i$) mean expectation value (determined by the $o$-th order CEM) and $N$ is the number of QEs. Similarly for $\langle\hat{\sigma}^{\tilde{k}}(\chi_i)\rangle_{\textbf{FQD}}$, where we utilized the FQD. The error definitions in this section provide the quantification of how reliable a given CEM solution is. 

\section{The good: Smaller error with higher order}\label{suse:thegood}
Model 1 in Sec.~\ref{suse.mod1} demonstrates the success of the CEM. Considering Eq.~(\ref{eq:error}), we define $\tilde{k}\in\{22,z,x\}$ and $\chi_i = \Gamma t_i$ with $t_i = i\cdot\delta t$, $i\in\{0,1,\dots,M\}$, $\delta t = T/M$ and $T=10$ is the overall evolution time. In Eq.~(\ref{eq:ham1}), we specifically write $\eta \in \mathbb{R}$ (constant drive), however, we assume that after time $\Gamma t_1 = \Gamma\cdot T/2 = 5$ the drive is turned off. Formally, we redefine $\eta\to\eta' = \eta\cdot(\Theta(\Gamma t) - \Theta(\Gamma (t - t_1))).$ We expect, that the system experiences excitation dynamics for $\Gamma t \leq 5$ and relaxes collectively into its ground state for $\Gamma t> 5$.

\begin{figure}[h!]
    \centering
    \includegraphics[width=0.99\linewidth]{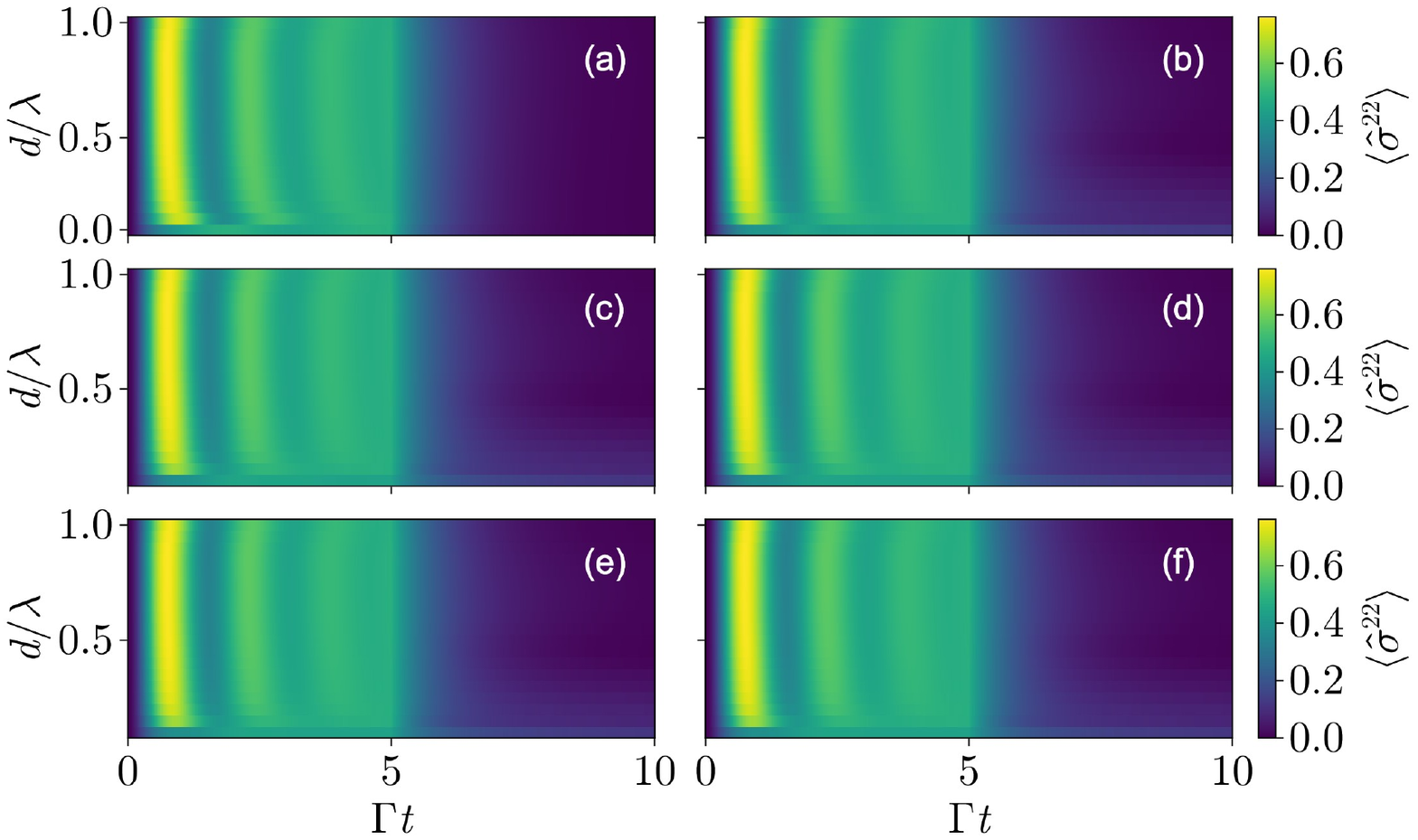}
    \caption{The mean populations $\langle\hat{\sigma}^{22}\rangle$ for $N = 5$, $\eta/\Gamma = 2$, $\Gamma t\in[0,10]$ and $d/\lambda\in\{0.1,0.15,0.2,\cdots,1.0\}$ in different CEM expansion orders $o\in\{1,2,3,4,5\}$ are visualized. With each plot ((a)-(e)), $o$ is increased, hence, plot (a) corresponds to $o=1$, plot (c) to $o = 3$ and plot (e) to $o=5$. In plot (f), the FQD is shown.}\label{fig.heat1}
\end{figure}
\begin{figure}[h!]
    \centering
    \includegraphics[width=0.5\textwidth]{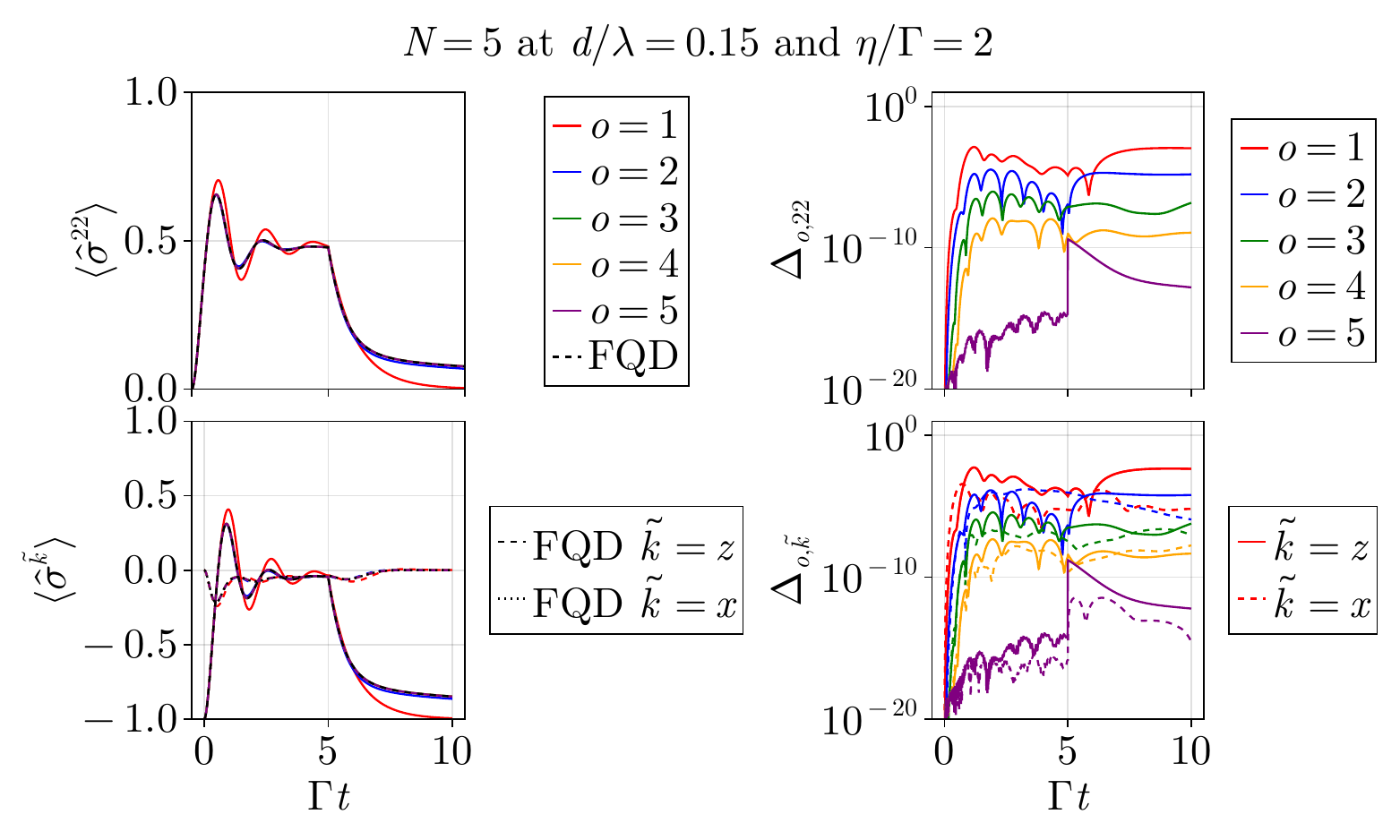}
    \caption{Comparison of the CEM in different orders $o\in\{1,2,3,4,5\}$ with the FQD of the expectation values $\langle\hat{\sigma}^{\tilde{k}}\rangle$ can be observed in the left plot column in sub-wavelength separation $d/\lambda = 0.15$. The right column depicts the corresponding SD $\Delta_{o,\tilde{k}}$.}\label{fig.case1}
\end{figure}
\begin{figure}[h!]
    \centering
    \includegraphics[width=0.5\textwidth]{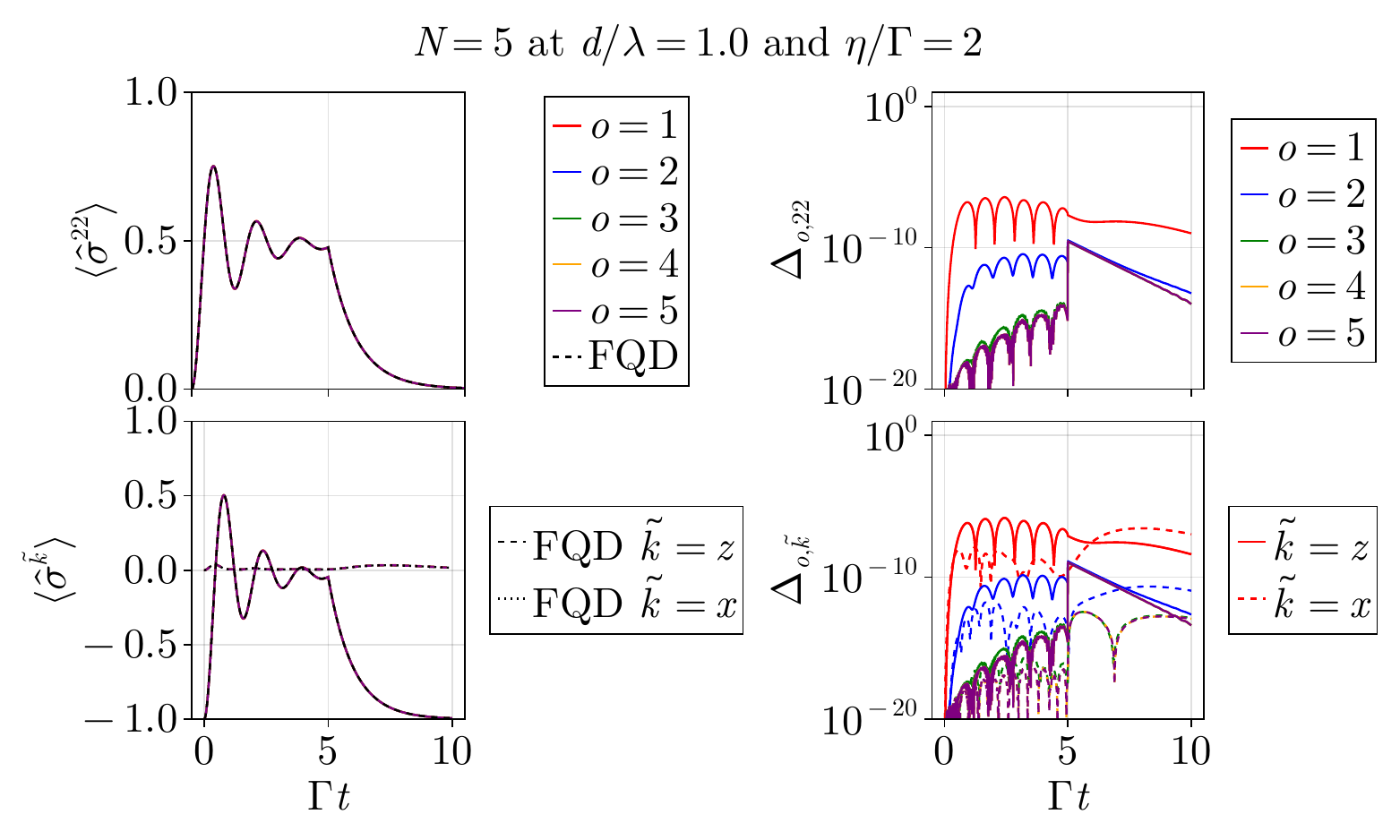}
    \caption{The CEM in different orders $o\in\{1,2,3,4,5\}$ is again compared with the FQD of the expectation values $\langle\hat{\sigma}^{\tilde{k}}\rangle$ in the left plot column with $d/\lambda = 1$. The right column visualizes the SD $\Delta_{o,\tilde{k}}$.}\label{fig.case2}
\end{figure}

In Fig.~\ref{fig.heat1}, we find the expected behavior. The drive $\eta$ induces excitation dynamics for $\Gamma t \leq  5$ and the system undergoes a relaxation process for $\Gamma t>5.$ The first-order CEM, i.e.\ $o = 1$, in plot (a) deviates slightly from higher-order expansions (see. plots ((b)-(e))) for small $d/\lambda$, nevertheless, for larger $d/\lambda$ the first-order CEM coincides with the plots ((b)-(f)). No visual differences can be observed between the solutions of higher-order expansions with $o\in\{2,3,4,5\}$ and the full quantum treatment in plot (f). Thus in Fig.~\ref{fig.heat1}, higher-order expansions can be used to physically describe the problem nearly correctly. In Fig.~\ref{fig.case1} ($d/\lambda = 0.15,~\eta/\Gamma = 2$) and Fig.~\ref{fig.case2} ($d/\lambda = 1,~\eta/\Gamma=2$), this is visualized by comparing the FQD solution with the CEM, where $o\in\{1,2,3,4,5\}.$ Here, the interpolation dependent errors, defined in Eq.~(\ref{eq:error1}), are shown as well.  

\begin{figure}[h!]
    \centering
    \includegraphics[width=0.5\textwidth]{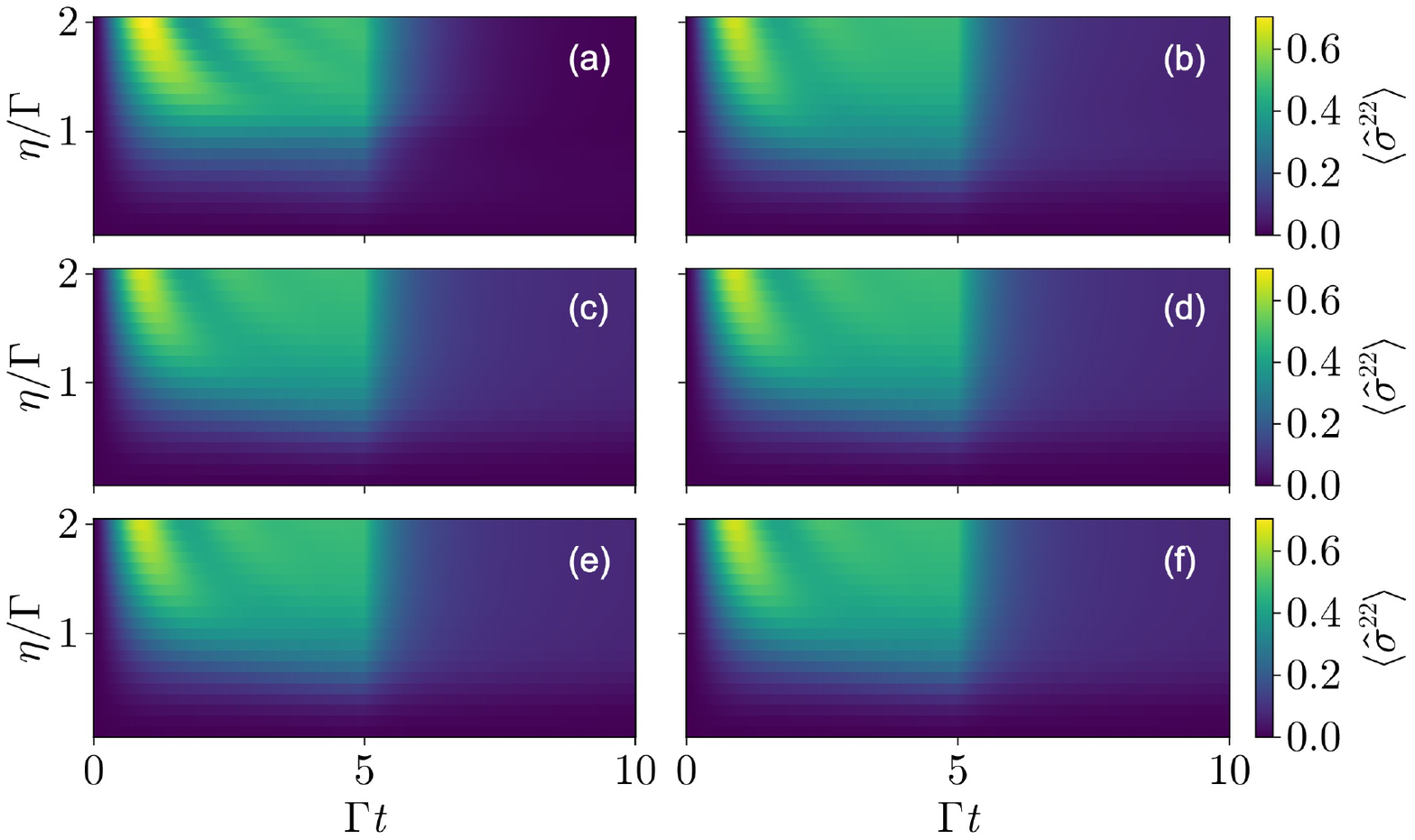}
    \caption{The mean populations $\langle\hat{\sigma}^{22}\rangle$ for $N = 5$, $d/\lambda = 0.15$, $\Gamma t\in[0,10]$ and $\eta/\Gamma\in\{0.1,0.2,\cdots,2.0\}$ in different orders $o\in\{1,2,3,4,5\}$ of CEM are visualized. With each plot ((a)-(e)), $o$ is again increased. Plot (a) corresponds to $ o=1$, plot (c) to  $ o = 3$ and plot (e) to $o=5$. In plot (f), the FQD is shown.}\label{fig.heat2}
\end{figure}

We find, indeed, that in case of deep sub-wavelength separation (strong dipole-dipole interaction), the first-order expansion is clearly distinguishable from higher-order expansions, introducing a relatively large SE $\Delta_{o,\tilde{k}}$. However, for larger distances the mean-field solution ($o = 1$) successively coincides with the higher-order solutions as well as the full quantum treatment. The higher-order expansions provide in both, Fig.~\ref{fig.case1} and Fig.~\ref{fig.case2}, the correct solutions with high precision. Similarly, we can also visualize the correctness of the CEM solutions by varying the drive strength $\eta/\Gamma$. This is shown in Fig.~\ref{fig.heat2}.

In case of Fig.~\ref{fig.heat2}, one can clearly observe a different pattern for the mean-field expansion. Especially for large $\eta/\Gamma$, the results differ from higher-order expansions as well as the full quantum treatment. As in Fig.~\ref{fig.heat1}, the higher-order expansions describe the problem sufficiently for all cases of $\eta/\Gamma$. 

\begin{figure}[h!]
    \centering
    \includegraphics[width=0.5\textwidth]{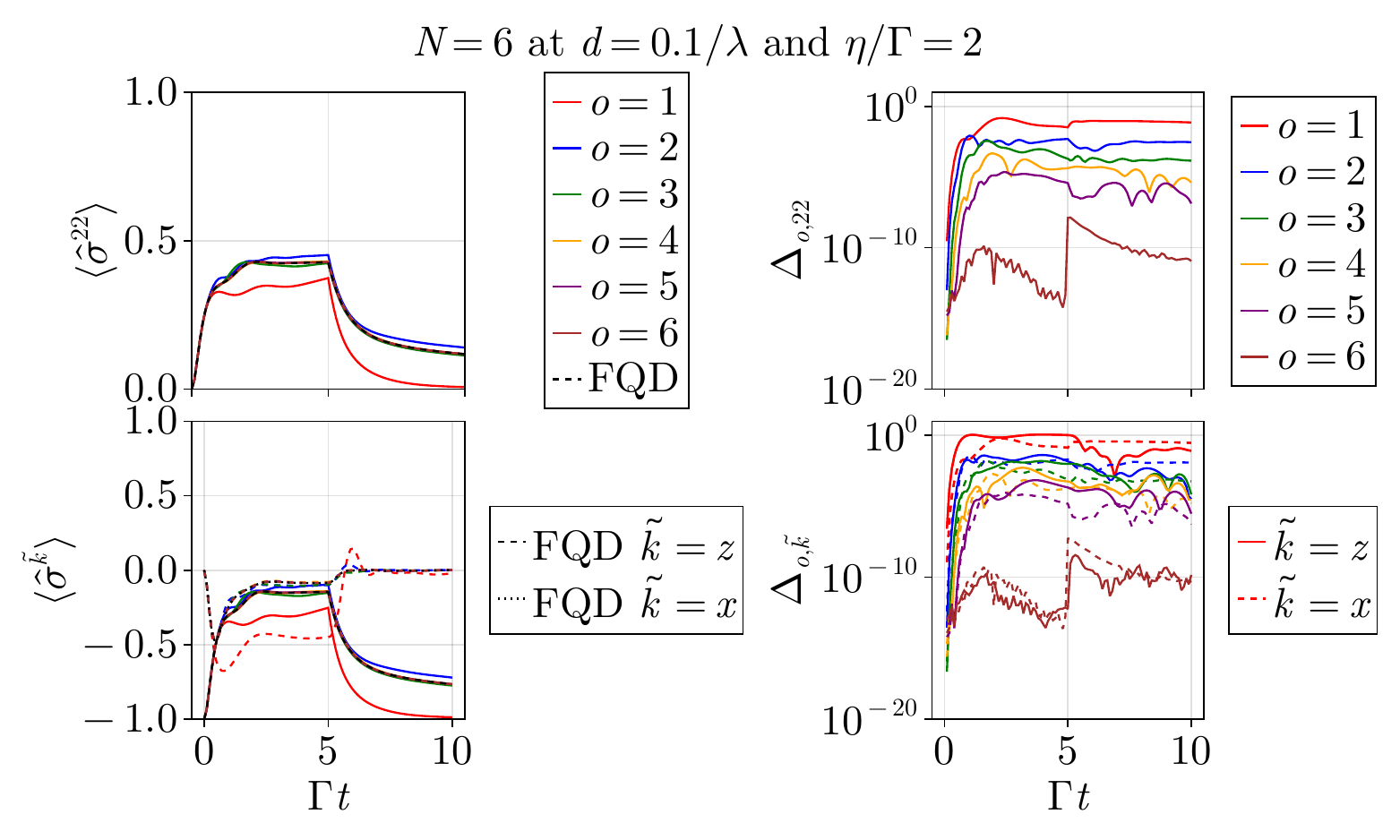}
    \caption{Comparison of the CEM for $N = 6$ in different orders $o\in\{1,2,3,4,5,6\}$ with the FQD of the expectation values $\langle\hat{\sigma}^{\tilde{k}}\rangle$ can again be observed in the left plot column. The right column depicts the corresponding SE $\Delta_{o,\tilde{k}}$.}\label{fig.case3}
\end{figure}

Increasing the number of QEs to $N = 6$ shows the same behavior as above. This is depicted in Fig.~\ref{fig.case3}.

\begin{table}[h!]
\caption{\label{tab1}
The values of the general error measure $\tilde{\Delta}_{o,\tilde{k}}$ are represented for $N = 6$, $d/\lambda = 0.1$, $\eta/\Gamma = 2$ and with $o\in\{1,2,\dots,6\}$ are presented here.}
\begin{ruledtabular}
\begin{tabular}{ccccc}
$o$ & $\tilde{\Delta}_{o,22}$ & $\tilde{\Delta}_{o,z}$ &  $\tilde{\Delta}_{o,x}$\\
\hline
$1$ & $1.983\cdot10^{-1}$ & $7.932\cdot10^{-1}$ & $1.370\cdot10^0$\\
$2$ & $7.920\cdot10^{-3}$ & $3.168\cdot10^{-2}$ & $3.669\cdot10^{-2}$ \\
$3$ & $1.512\cdot10^{-3}$ & $6.051\cdot10^{-3}$ & $1.430\cdot10^{-2}$ \\
$4$ & $1.735\cdot10^{-4}$ & $6.941\cdot10^{-4}$ & $2.457\cdot10^{-3}$ \\
$5$ & $1.405\cdot10^{-5}$ & $5.619\cdot10^{-5}$ & $4.106\cdot10^{-4}$ \\
$6$ & $1.931\cdot10^{-9}$ & $7.723\cdot10^{-9}$ & $5.764\cdot10^{-10}$
\end{tabular}
\end{ruledtabular}
\end{table}

As a qualitative statement for the chain of , e.g., $N = 6$ dipole-interacting two-level QEs in Fig.~\ref{fig.case3}, the first-order expansion has a computation time of few seconds, providing much faster useful results as the fifth-order expansion with more than $15$h runtime. However, the price paid is the decrease of the physical accuracy, schematically shown in Fig.~\ref{fig.1}. Since the second-order expansion already gives solutions of high precision with runtime up to $1$~min, this is a good choice for this problem. The adjusted general error measure $\tilde{\Delta}_{o,\tilde{k}}$ from Fig.~\ref{fig.case3}, defined in Eq.~(\ref{eq:error}), demonstrates this further in Tab.~\ref{tab1}. The values in Tab.~\ref{tab1} and the SE plots in Fig.~\ref{fig.case1}, Fig.~\ref{fig.case2} and Fig.~\ref{fig.case3} are used to label model 1 as "The Good". With increase of the expansion order, we gain accuracy, however, we are charged with computation time.

\section{The Bad: Higher Orders Introduce Artifacts}\label{suse:thebad}

\begin{figure}[h!]
    \centering
    \includegraphics[clip, trim=0.0cm 4.5cm 0.0cm 4.8cm, width=1.0\linewidth]{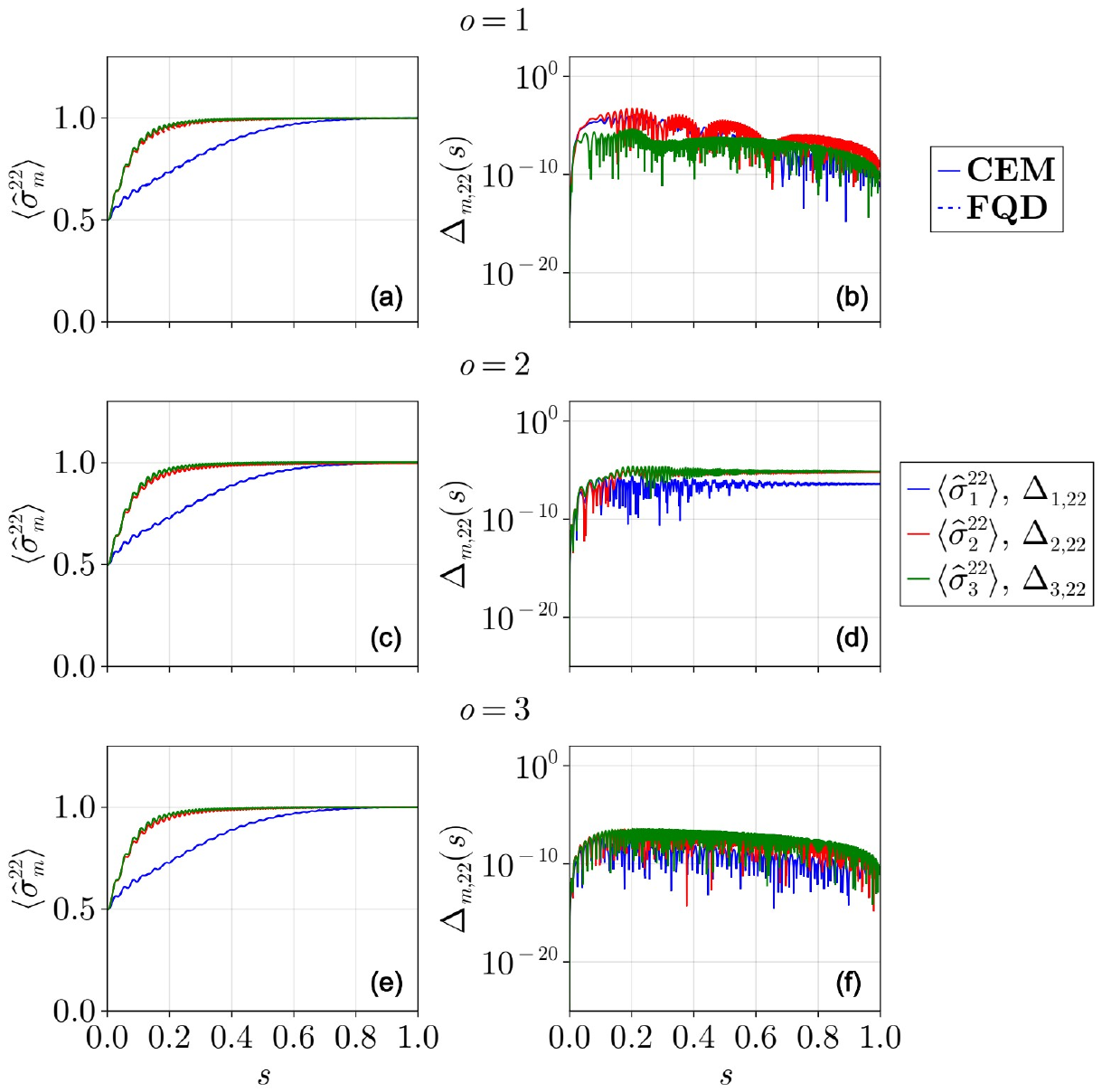}
    \caption{Comparing the CEM solutions of the bi-prime factorization problem for $\omega = 21 = 7[\textcolor{red}{1}\textcolor{blue}{1}1]\times[\textcolor{newgreen}{1}1]$ in different orders $o\in\{1,2,3\}$ together with the FQD determined by solving the Heisenberg EOM directly (left plot column). The correct bit digit sequence is acquired after the adiabatic sweep, i.e., $[\textcolor{blue}{a_1=1},\textcolor{red}{a_2=1},\textcolor{newgreen}{b_1=1}] = [\textcolor{blue}{\langle \hat{a}_1(1)\rangle},\textcolor{red}{\langle \hat{a}_2(1)\rangle},\textcolor{newgreen}{\langle \hat{b}_1(1)\rangle}]$, keeping in mind $\hat{a}_i = \hat{\sigma}^{22}_i$ and $\hat{b}_j = \hat{\sigma}_{k+j}^{22}$. The SD between CEM and FQD $\Delta_{o,\tilde{k},m}(s_i)$ are presented in the right plot column, where $\tilde{k}\in\{22\}$ and $T = 10$ in arbitrary time units.}\label{fig.bad}
\end{figure}

We switch to the in Sec.~\ref{suse.mod2} introduced notation, that the particle number is $n = N$. Including higher-order correlations can improve the physical description of a considered system as demonstrated in the previous Sec.~\ref{suse:thegood}. However, numerous cases are known, where the CEM fails to provide better solutions with increasing expansion order $o = 1 \to n$ \cite{PhysRevB.105.224305,PhysRevResearch.5.033148,photonics12100996}. Depending on the problem, the neglection of certain quantum correlations can imply the construction of an unstable ODE system, causing non-convergent or wrong results due to numerical artifacts \cite{photonics12100996} ("The Bad").
The bi-prime factorization problem, discussed in Sec.~\ref{suse.mod2}, turns out to describe such a "Bad" problem by applying the CEM to it. First, it produces mean-field ($o=1$) as well as $n$-order expansion solutions coinciding with the statements in Fig.~\ref{fig.1}, matching the expected behavior ("The Good"-features), i.e., the mean-field solution is correct but with relatively large error and the $n$-order expansion is the FQD (determined by the Heisenberg EOM in Eq.~(\ref{eq.Heisenberg})). This can be observed in Fig.~\ref{fig.bad} for the toy model $\omega = 21 = 7[\textcolor{red}{1}\textcolor{blue}{1}1]\times[\textcolor{newgreen}{1}1]$, where three qubits are required ($n = k + l = 3$). As it was mentioned in Sec.~\ref{suse.mod2}, excluding the second minimum ensures a non-degenerate ground state energy, thus, we choose $k = 2$ and $l = 1$.

In Fig.~\ref{fig.bad}(a), no visual deviation between mean-field and FQD solution can be observed, hence, the final state $\psi(s=1)$ is already close enough to the wanted ground state of $\hat{H}_{\text{p}}$, yielding the correct expectation values which solve the bi-prime problem $\omega = 21.$ We point out that we neglect here all quantum correlations, i.e.\ $\langle \hat{O}_1\cdots\hat{O}_n\rangle = \langle \hat{O}_1\rangle \cdots\langle\hat{O}_n\rangle$. One might think that the mean-field approach can be used to efficiently solve the bi-prime problem for arbitrary integers. However, further investigations indicate that for growing qubit systems a much larger sweeping time $T$ is required to obtain the correct classical bit-digits, slowing down the computation process rapidly and eventually making first-order CEM impractical. This means the efficient determination of the correct classical bit-digit sequence is bound to $T$, indicating a possible Quantum Advantage. Fig.~\ref{fig.bad}(c) visualizes that increasing the order to $o = 2$ does not provide a better solution. In fact, comparing the SE time-evolutions of the second-order (plot (d)) with the first-order (plot (b)) expansion shows, that the mean-field approximation produces even an numerically better bit-digit sequence than the solution including bipartite entanglement terms, i.e., $\langle \hat{O}_i\hat{O}_j\rangle \neq \langle \hat{O}_i\rangle \langle\hat{O}_j\rangle$. From Fig.~\ref{fig.bad}~((e)-(f)), it is clear that FQD is obtained in third-order expansion. In Tab.~\ref{tab2}, we demonstrate that the general error measures of $o = 1$ and $o = 2$ are in the same regime, whereas $o = 3$ is between one and two orders of magnitude smaller. Due to the above discussion we find that increasing the CEM expansion order does not necessarily yield better results for the trade of increased computation time as debated in Fig.~\ref{fig.1} and Sec.~\ref{suse:thegood}, revealing the "Bad" nature of the bi-prime factorization problem.

\begin{table}[h!]
\caption{\label{tab2}
The values of the general error measure $\tilde{\Delta}_{o,\tilde{k},m}$ corresponding to the bi-prime problem $\omega = 21$.}
\begin{ruledtabular}
\begin{tabular}{ccccc}
$o$ & $\tilde{\Delta}_{o,22,1}$ & $\tilde{\Delta}_{o,22,2}$ & $\tilde{\Delta}_{o,22,3}$ \\\hline
$1$ & $5.36\cdot10^{-3}$                 & $2.62\cdot10^{-2}$                 & $3.24\cdot10^{-4}$                 \\
$2$ & $3.32\cdot10^{-4}$        & $4.88\cdot10^{-3}$                & $6.71\cdot10^{-3}$        \\
$3$ & $2.94\cdot10^{-6}$     & $4.86\cdot10^{-5}$     & $8.35\cdot10^{-5}$     
\end{tabular}
\end{ruledtabular}
\end{table}

\section{The Ugly: Erratic Time Evolution}\label{suse:theugly}
\begin{figure}[h!]
    \centering
    \includegraphics[clip, trim=0.0cm 4.5cm 0.0cm 4.8cm, width=1.0\linewidth]{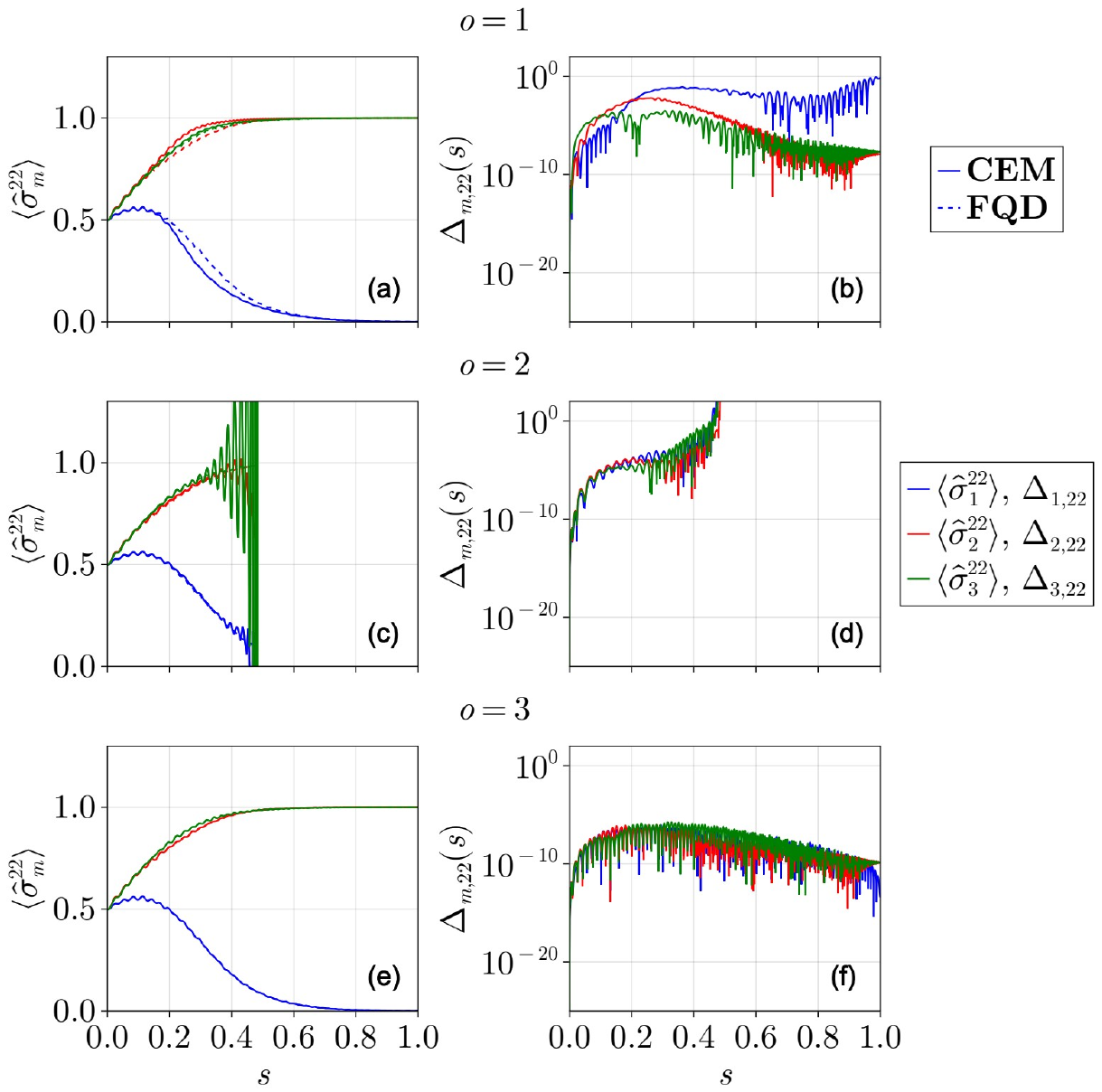}
    \caption{Comparing the CEM solutions of the bi-prime factorization problem for $\omega = 15 = 5[\textcolor{red}{1}\textcolor{blue}{0}1]\times 3 [\textcolor{newgreen}{1}1]$ in different orders $o\in\{1,2,3\}$ with the FQD (left plot column). The correct bit digit sequence is obtained after the adiabatic transfer, i.e., $[\textcolor{blue}{a_1=0},\textcolor{red}{a_2=1},\textcolor{newgreen}{b_1=1}] = [\textcolor{blue}{\langle \hat{a}_1(1)\rangle},\textcolor{red}{\langle \hat{a}_2(1)\rangle},\textcolor{newgreen}{\langle \hat{b}_1(1)\rangle}]$, once again keeping in mind $\hat{a}_i = \hat{\sigma}^{22}_i$ and $\hat{b}_j = \hat{\sigma}_{k+j}^{22}$. The SD between CEM and FQD $\Delta_{o,\tilde{k},m}(s_i)$ are shown in the right plot column, where $\tilde{k}\in\{22\}$ and $T = 10$ in arbitrary time units. }\label{fig.ugly}
\end{figure}

Considering different target numbers $\omega$ in the bi-prime factorization problem not only shows that "The Bad" property is fulfilled, but one finds that "The Ugly" is present, i.e.\ the resulting equation already in second order behave wildly and exhibit uncontrolled divergent behavior. While at meanfield (first order) level we get reasonably well converging results up to about 10 effective qubits, things at the second order already change for much lower numbers to factorize. Even trying to factorize a number as low as only 15 requiring only 3 qubits to represent the factors, the second order equations produce nonphysical fast oscillations and divergencies and partly even prevent numerical integration at all. This can be seen in Fig.~\ref{fig.ugly}, where we depict the time evolutions of the first spin moments for this case ($\omega = 15$) calculated via an expansion first, second and third order. While the first order again behaves reasonable and the third order represents the exact results as it should, the second order solutions are simply erratic. Such an erratic behavior from a 'harmless'  set of low order polynomial differential equations is not only unexpected but also raises the question: what evokes this behavior and do the solutions remain physical and numerical instabilities are the cause of this or do these divergences appear due to truncation instabilities? 

 As argued in Sec.~\ref{suse:thebad}, the mean-field solution in Fig.~\ref{fig.ugly}(a) yields the correct classical bit-digit sequence, now with visual deviations from the FQD (Heisenberg EOM), and the FQD is determined in third-order expansion in Fig.~\ref{fig.ugly}(e). The interesting case can be observed in Fig.~\ref{fig.ugly}(c). The CEM not only fails (worse solution than first-order expansion) ("The Bad"), but also produces nonphysical divergent results ("The Ugly"). The bit-digits cannot be extracted, leaving us with no useful information about the considered problem.
 
A more thorough investigation with different state-of-the-art explicit solvers did not lift these divergences and did not provide any solutions beyond critical time. For $\omega = 15$, we found implicit solvers, where the divergences vanished. However, investigation of larger integers revealed that even said solvers cannot mitigate the instabilities in general. In order to obtain insights about the internal ODE system's structure and behavior, we have considered further methods.
 
A linear stability analysis approach was applied. Although the time-evolution eigenvalues of the ODE system's Jacobian have revealed the divergent behavior in the form of oscillations, a decisive criterion for stiffness could not be found. A more advanced stability analysis approach could be interesting, however, this is beyond the scope of this work.

We have introduced a dissipative environment, but discussed problems remain. To this moment we have not yet identified the origin of these suddenly appearing strange divergences in quite a few specific cases. Indeed, it seems that the three- and four-body interaction-terms appearing in $\hat{H}_{\text{p}}$ induce correlations and introduce internal oscillations, which are canceled by the highest order correlations only.
 
In general such detrimental oscillations appear already in the equations for the second-order momenta, which gets even worse in higher orders up to the next until the last unstable order $o = n-1$ is reached. Therefore, we only obtain well-behaved solutions (finite values between $0$ and $1$) for $o = 1$ (no correlations) and $o = n$ (all correlations). Not only did we observe unstable results for all investigated integers $\omega\in\{51,57,63,85,91,93\}$ with $n > 4$, but also for $\omega = 15~(n=3)$ in Fig.~\ref{fig.ugly} and $\omega = 33~(n = 4)$. Exceptional cases with finite values are $\omega =21~(n = 3)$ in Fig.~\ref{fig.bad} and $\omega = 39~(n = 4)$. This can be seen as an indication that the bi-prime factorization problem, which is solved via utilization of the QOP-Hamiltonian in Eq.(~\ref{eq.Hp}), is only reasonably solvable at a classical mean field or a full quantum level. Higher order corrections to the mean field only make the numerics slower and the approximation worse, so these are simply useless. The general error measure values, presented in Tab.~\ref{tab3}, embody the above discussion of the nonphysical divergent ("Ugly") nature of the bi-prime factorization problem. 

\begin{table}[h!]
\caption{\label{tab3}
The values of the general error measure $\tilde{\Delta}_{o,\tilde{k},m}$ corresponding to the bi-prime problem $\omega = 15$.}
\begin{ruledtabular}
\begin{tabular}{ccccc}
$o$ & $\tilde{\Delta}_{o,22,1}$ & $\tilde{\Delta}_{o,22,2}$ & $\tilde{\Delta}_{o,22,3}$ \\\hline
$1$ & $0.96$                 & $0.65$                 & $0.03$                 \\
$2$ & $1.05\cdot10^6$        & $1.87\cdot10^4$               & $2.02\cdot10^8$        \\
$3$ & $9.60\cdot10^{-5}$     & $7.19\cdot10^{-5}$     & $1.9\cdot10^{-4}$     
\end{tabular}
\end{ruledtabular}
\end{table}

\section{Conclusions and Outlook}
At the examples of finite coupled spin systems with two different types of interactions, which can be numerically solved in full,  we have exhibited the usefulness and uselessness of a moment based expansion of their quantum dynamics in the form of the known cumulant expansion. An extensive quantitative analysis of the accuracy and convergence properties of various orders of the cumulant expansion in comparison with the full (exact) quantum dynamics reveals expected as well as puzzling behavior.

While for binary and not too strong interactions one observes the expected convergence of the approximate results to the full numerical solution with increasing order ("The Good"), in the case of higher order interaction terms in the Hamiltonian one often finds cases, where higher orders even agree less with the full solution than lower orders ("The Bad"). In several supposedly rather simple cases, we even encounter completely uncontrolled oscillatory dynamics, far from any physically realistic or sensible solution ("The Ugly"). 

Our intent is to spark a discussion about the applicability of the cumulant expansion with the ultimate goal of finding rigorous a priori criteria for its application~\cite{kusmierek2023higher}.

\begin{acknowledgments}
We acknowledge early contributions of Maxim Laborenz in the dipole-dipole example (see: \textsl{https://ulb-dok.uibk.ac.at/ulbtirolhs/content/structure/9603428}). We thank Klemens Hammerer for helpful comments and discussions and Alexander Ostermann und Lukas Einkemmer for valuable advice on the numerical simulations. These were performed using the open-source frameworks QuantumOptics.jl~\cite{kramer2018quantumoptics} and QuantumCumulants.jls~\cite{plankensteiner2022quantumcumulants}.

This research was funded in whole or in part by the Austrian Science Fund (FWF) projects 10.55776/FG5 (Forschungsgruppe FG 5) and the quantA cluster of excellence 10.55776/COE1.
\end{acknowledgments}

\bibliography{refs}

\end{document}